\documentclass[preprint,showpacs,showkeys,preprintnumbers,aps]{revtex4}

\usepackage{graphicx}% Include figure files
\usepackage{float}% Include hold figures
\usepackage{amsmath}

\begin{document}

% Title of the article
\title{Fabrication and characterization of semiconducting half Heusler YPtSb thin films}

% Abbreviated title for the page headers
%\titlerunning{Short title }

% Authors
\author{Rong~Shan}
\affiliation{Institut f\"ur Anorganische und Analytische Chemie,
Johannes Gutenberg - Universit\"at, 55099 Mainz, Germany.}
\author{Enrique~V.~Vilanova}
\affiliation{Institut f\"ur Physik,
Johannes Gutenberg - Universit\"at, 55099 Mainz, Germany.}
\author{Juan~Qin}
\affiliation{Institut f\"ur Anorganische und Analytische Chemie,
Johannes Gutenberg - Universit\"at, 55099 Mainz, Germany.}
\author{Frederick Casper}
\affiliation{Institut f\"ur Anorganische und Analytische Chemie,
Johannes Gutenberg - Universit\"at, 55099 Mainz, Germany.}
\author{Gerhard~H.~Fecher}
\affiliation{Max-Planck-Institute f\"ur Chemische Physik fester Stoffe, N\"othnitzer Str.~40, 01187 Dresden, Germany.}
\author{Gerhard~Jakob}
\affiliation{Institut f\"ur Physik,
Johannes Gutenberg - Universit\"at, 55099 Mainz, Germany.}
\author{Claudia~Felser}
\affiliation{Max-Planck-Institute f\"ur Chemische Physik fester Stoffe, N\"othnitzer Str.~40, 01187 Dresden, Germany.}
\email{felser@cpfs.mpg.de}


\begin{abstract}
The semiconducting half-Heusler compound YPtSb has been predicted to convert into a topological insulator under the application of an appropriate degree of strain. In this study, \textit{p}-type semiconducting YPtSb thin films were prepared by magnetron co-sputtering, using a specially designed target. YPtSb thin films grown on MgO (100) substrates at $600^\circ$C showed a textured structure with the (111) plane parallel to the (001) plane
of MgO. Electrical measurements showed that the resistivity of the YPtSb films decreases with increasing temperature, indicating semiconductor-like behavior. The carrier density was as high as $1.15\times 10^{21}$~cm$^{-3}$ at 300~K. The band gap of the YPtSb thin films was around 0.1-0.15~eV, which was in good agreement with the theoretical prediction and the value measured for bulk YPtSb.
\end{abstract}

\maketitle   % please do not remove

\section{Introduction}

Spin injection and spin transportation are essential issues to be addressed in spintronics device applications. In recent years, a topological insulator (TI) has been predicted to behave as an ideal spin medium in which spin-up and spin-down carriers flow in separate channels without spin and energy dissipation.~\cite{Kane2005,Qi2008,Bernevig2006,Koenig2007,Chen2009} Theoretically, the spin diffusion length in a TI is unlimited, so that TIs would find promising applications in next-generation spintronics devices. Up to now, experimental research on topological insulators has focused on HgTe/CdTe quantum wells and Bi$_{2}$Te$_{3}$ series.~\cite{Koenig2007,Chen2009} Many ternary half-Heusler compounds with 18 valence electrons function as semiconductors; addition of the atoms of a third element to the Wyckoff \textit{4b} positions in the $C1_b$ structure may result in an extended lattice similar to that of zincblende, a binary semiconductor, ~\cite{CAS2012,HAA02,CAS08} thus making the electronic structures of these two types of semiconductors similar. In 2010, Chadov et al. predicted that numerous half-Heusler semiconductors such as LuPtBi, LuPdBi, and LaPtBi show band inversion similar to that of HgTe, because of the strong spin-orbit coupling of the constituent heavy elements. ~\cite{Chadov2010} The near-zero-gap characteristic of some of these compounds has also been demonstrated experimentally on the basis of their bulk properties. ~\cite{OUA2011,OUA2011b,SHA12,SHE12} The first thin film of LaPtBi was grown on a YAlO$_3$ substrate by three-source magnetron co-sputtering.~\cite{MIY2012}
The TI state could also be induced in half-Heusler semiconductors comprising lighter elements, such as YPtSb and YPdBi, by applying a certain degree of strain.~\cite{Chadov2010} 
In comparison with HgTe, most half-Heusler TIs have stable chemical properties, and films of these compounds can be easily prepared by sputtering, a commonly used technique for device fabrication. The half-Heusler YPtSb is close to the border between the trivial semiconductors and topological insulators, and it has a narrow gap of around 0.16~eV.~\cite{Oestreich2003} Every 0.01~{\AA} increase in the lattice constant along the c axis has been calculated to result in about a 20~meV energy difference between the inverted bands. In the case of the half-Heusler YPtSb, the lattice constant is around 6.56~{\AA}, and hence, even a 1\% variation in the lattice may trigger a trivial-semiconductor-to-TI phase change, while opening the band gap near the Dirac point. 

% Figure 2 %%%%%%%%%%%%%%%%%%%%%%%%%%%%%%%%%%%%%%%%%%%%%%%%%%%%%%%%%
\begin{figure}[ht]
   \includegraphics[width=\linewidth]{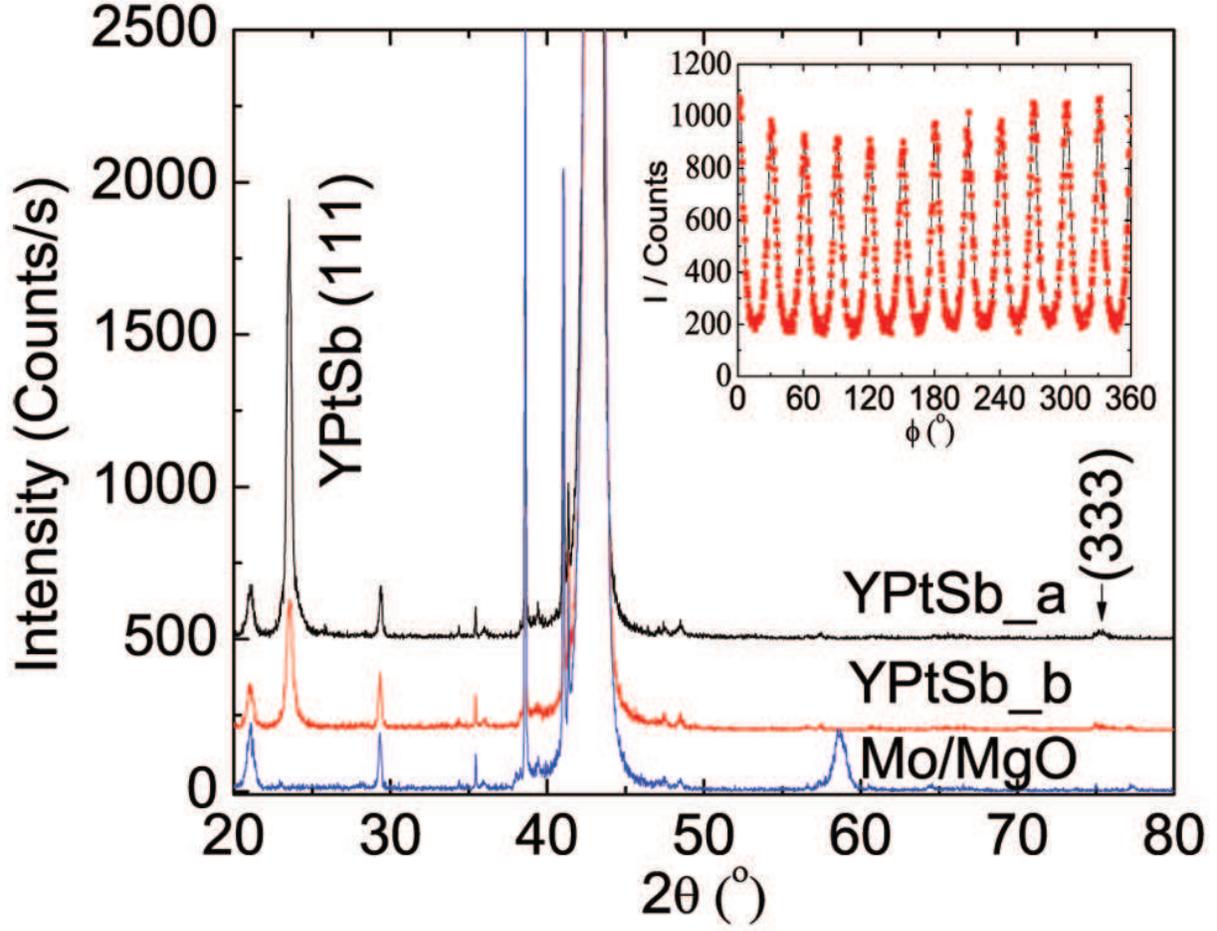}
   \caption{ (Color online) X-ray diffraction patterns of YPtSb thin films grown on MgO (100) substrates at 3~Pa and $800^\circ$C (YPtSb\_a) and at 0.3~Pa and $600^\circ$C (YPtSb\_b). The pattern of the Mo thin film deposited on MgO(100) shows the background peaks. The inset shows the $\phi$ scan of the (220) peaks. }
\label{fig:str}
\end{figure}
%%%%%%%%%%%%%%%%%%%%%%%%%%%%%%%%%%%%%%%%%%%%%%%%%%%%%%%%%%%%%%%%%%%%

\section{Experimental Details}

MgO(100)/YPtSb (100~nm) samples were prepared by DC magnetron sputtering at different substrate temperatures n Ar atmosphere, using a chamber with a base pressure lower than $5\times 10^{-7}$~Pa. The composition of the YPtSb thin films was analyzed by energy-dispersive X-ray spectroscopy (EDX). The Fourier transform infrared spectroscopy (FTIR) spectra were obtained using a Nicolet 730 spectrometer equipped with a diamond attenuated total reflection (ATR) crystal and a liquid N$_{2}$-cooled mercury-cadmium-telluride detector. Electrical resistance and Hall effect measurements were performed using a physical property measurement system (Quantum Design, PPMS), under a magnetic field of up to 7~T, using the six-lead method. For accurate measurement of the Hall coefficient, Hall resistivity measurements were performed by sweeping the magnetic field at a fixed temperature.

% Figure 3 %%%%%%%%%%%%%%%%%%%%%%%%%%%%%%%%%%%%%%%%%%%%%%%%%%%%%%%%%
\begin{figure}[ht]
   \includegraphics[width=\linewidth]{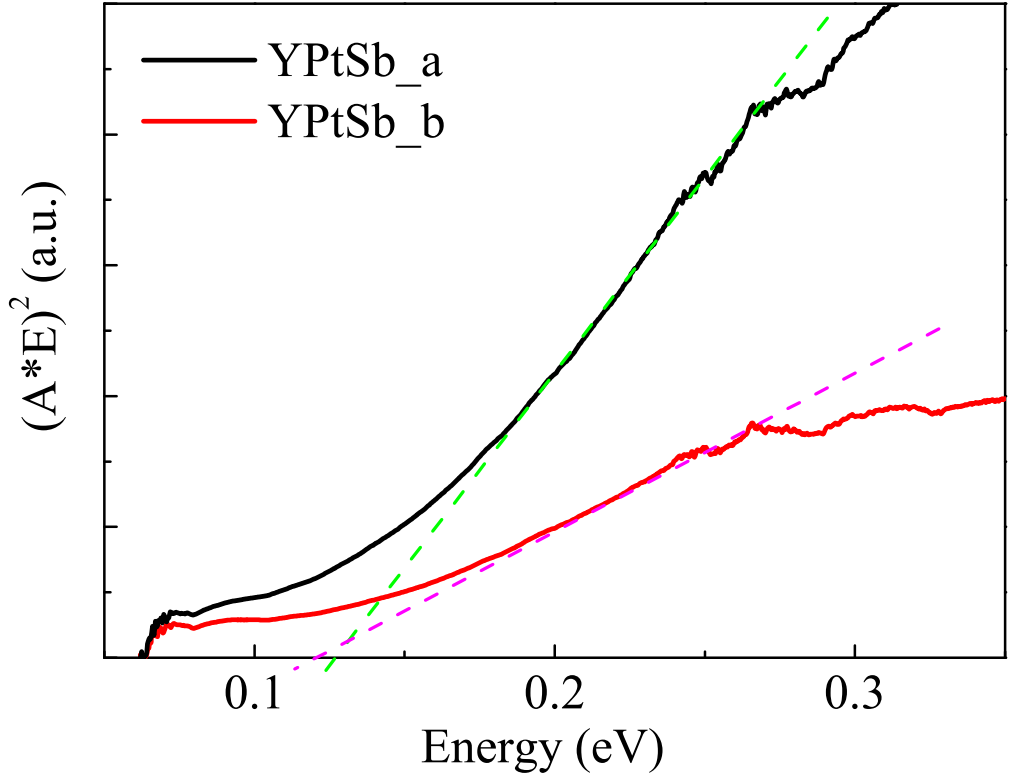}
   \caption{(Color online) Tauc plot to resolve optical band gap obtained from ATR-FTIR spectra of YPtSb\_a and YPtSb\_b thin films. The two dotted lines indicate the linear fitting to the data.}
\label{fig:ir}
\end{figure}
%%%%%%%%%%%%%%%%%%%%%%%%%%%%%%%%%%%%%%%%%%%%%%%%%%%%%%%%%%%%%%%%%%%%

\section{Results and Discussion}
A $cake$ target consisting of sector segments of the different metals was used to obtain stoichiometric films. The shape of the small pieces was fixed as a sector so that the deposition rate is unchanged during prolonged sputtering. The composition of the compounds was tuned by varying the size of each piece. Analysis of two different regions on each sample surface revealed that the composition ratios of the films are uniform and close to the stoichiometric ratio of 1:1:1 after four rounds of tuning. Figure ~\ref{fig:str} displays the X-ray diffraction (XRD) patterns of the YPtSb thin films deposited on MgO (100) substrates (YPtSb\_a) at 3~Pa and $800^\circ$C (the heater temperature, and the same hereinafter) and (YPtSb\_b) at 0.3~Pa and $600^\circ$C. The YPtSb film quality was independent of the Ar pressure. For these films, only (111) peaks
of YPtSb are observed. Although both YPtSb ($C1_b$, a $\approx6.53$~{\AA}) and MgO (B1, a $\approx4.21$~{\AA}) have a cubic structure, the large lattice mismatch between YPtSb (100) and MgO (100), around 55\% along cubic axis and around 10\% for 45° rotated growth, was not conducive for epitaxial growth. YPtSb films  grew with the dense (111) plane on MgO (100) substrate. There exists even an in-plane texture. The $\phi$ scan of the (220) peak exhibits 12 peaks, as shown in the inset of Fig.~\ref{fig:str}.
These twelve peaks result from four symmetry equivalent crystalline variants having their [1-10] direction 
parallel to one of the cubic in-plane axis of MgO (i.e. [010], [0-10], [001], [00-1] directions).

Figure~\ref{fig:ir} shows the ATR-FTIR spectra of the YPtSb\_a and YPtSb\_b thin films. Because the spectral intensity decreases at longer wavelengths of ATR absorbance, the raw data were corrected by using the onboard software on the instrument. The absorption coefficient was calculated by dividing the absorbance by the film thickness. The band gaps of both the samples were between 0.1~eV and 0.15~eV, which were in good agreement with that of bulk YPtSb, as revealed by high-temperature resistivity measurements.~\cite{Oestreich2003}

\begin{figure}[ht]
   \includegraphics[width=0.7\linewidth]{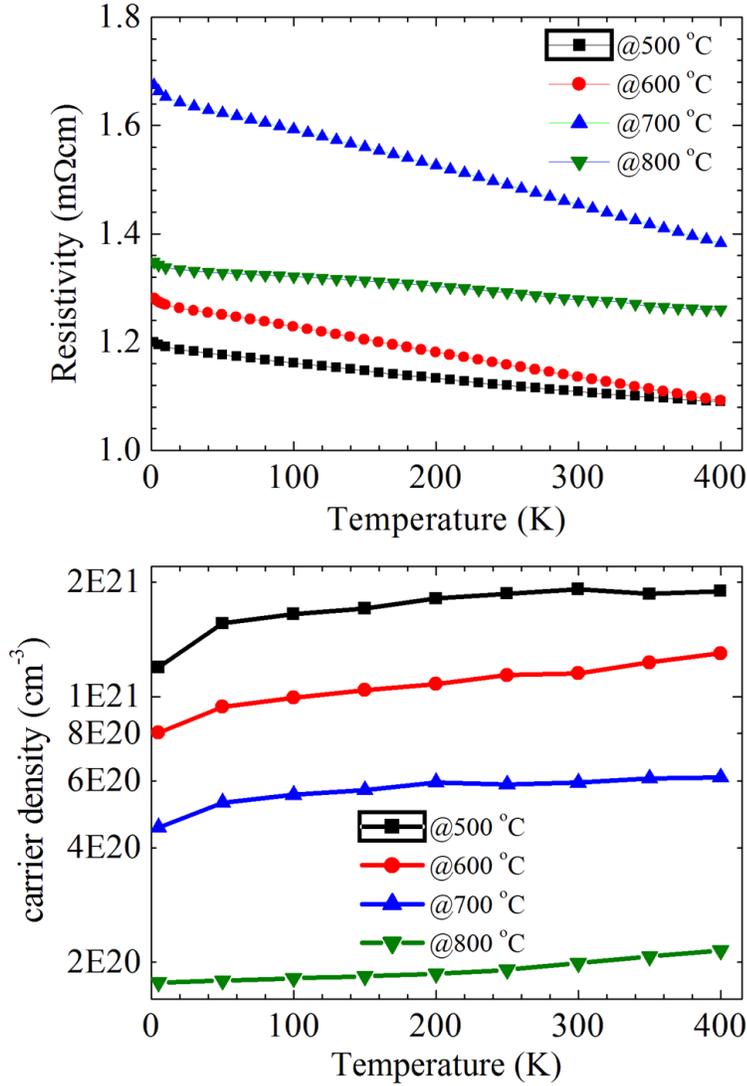}
   \caption{(Color online) Electrical transport properties of YPtSb thin films deposited at different temperatures.
    (a) Temperature dependence of resistivity; (b) Temperature dependence of carrier density. }
\label{fig:etp}
\end{figure}

The temperature dependence of the electrical resistivity and carrier density is presented in Figs.~\ref{fig:etp}(a) and (b), respectively. The decreasing resistivity with increasing temperature indicated semiconductor-like behavior in all the YPtSb thin films. Hall Effect measurements showed that the YPtSb thin films were p-type conductors with a high carrier density on the order of $10^{20} - 10^{21}$~cm$^{-3}$, which corresponded to the weak semiconducting behavior indicated by the resistivity curves. The weak temperature dependence of resistivity has also been observed in bulk YPtSb and other half-Heusler semiconductors~\cite{SHA12b}, a trend markedly different from that in the case of conventional semiconductor materials such as Si.~\cite{OUA2011,Xia2001,Uher1999,Ouardi2010} As seen in Fig.~\ref{fig:etp}(a) the resistivity increased from $500^\circ$C deposition temperature to $700^\circ$C, mainly because of the decreasing carrier density with an increase in deposition temperature (Fig.~\ref{fig:etp}(b)). The resistivity of the film deposited at $800^\circ$C was lower than that of the film deposited at $700^\circ$C, because the strongly enhanced carrier mobility counteracted the influence of the decreased carrier density.
% Figure 4 %%%%%%%%%%%%%%%%%%%%%%%%%%%%%%%%%%%%%%%%%%%%%%%%%%%%%%%%%
%%%%%%%%%%%%%%%%%%%%%%%%%%%%%%%%%%%%%%%%%%%%%%%%%%%%%%%%%%%%%%%%%%%%
\section{Summary and Conclusions}
Semiconducting half-Heusler YPtSb thin films were grown on MgO (100) substrates by using a target designed specially. Preliminary characterization proved that the YPtSb films are p-type conductors with a narrow band gap, as was also predicted by theoretical studies and reported for bulk YPtSb. High annealing temperatures could help in reducing carrier density and enhancing carrier mobility to afford high-quality thin films. Moreover, epitaxial growth of YPtSb films was also obtained on Ta/Mo buffered MgO (100) substrates, but the expected strain was not observed at the buffer/YPtSb interface since the film quality required further improvement.  

\section{acknowledgement}
The authors wish to thank S.S.P. Parkin for the valuable support and fruitful discussions and A. Dion for the technical support.  Financial support by the Deutsche Forschungsgemeinschaft (projects TP 4.5-A "`Heusler materials for new spin transport applications "` and 1.2-A "`The theory of electronic and magnetic structure of advanced spintronic materials with emphasis on spectroscopy and properties"' in research unit FOR 1464 'ASPIMATT') and  by the ERC Advanced Grant (291472) "`Idea Heusler"' is also gratefully acknowledged.

% Use the following code if you wish to generate your bibliography with BibTeX;
% replace the string "pss-demo" below with the name(s) of
% the BibTeX data base(s) you want to use.
% The resulting bibliography-output (the content of the .bbl file)
% must be pasted back into this file before submission.
% Please also include your BibTeX data base file(s) in your submission
% so that we can re-run BibTeX if necessary.
%
%\bibliographystyle{pss}
%\bibliography{YPtSb}

\begin{thebibliography}{[10]}

\bibitem{Kane2005}% article
 C.\,L. Kane and  E.\,J. Mele,
 Phys Rev Lett \textbf{95}, 146802 (2005).


\bibitem{Qi2008}% article
 X.\,L. Qi,  T.~Hughes,  and  S.\,C. Zhang,
 Phys.Rev. \textbf{B78}, 195424 (2008).


\bibitem{Bernevig2006}% article
 B.\,A. Bernevig,  T.\,L. Hughes,  and  S.\,C.
  Zhang,
 Science \textbf{314}(5806), 1757 (2006).


\bibitem{Koenig2007}% article
 M.~K\"onig,  S.~Wiedmann,  C.~Br\"une,
  A.~Roth,  H.~Buhmann,  L.\,W. Molenkamp,
  X.\,L. Qi,  and  S.\,C. Zhang,
 Science \textbf{318}(5851), 766 (2007).


\bibitem{Chen2009}% article
 Y.\,L. Chen,  J.\,G. Analytis,  J.\,H. Chu,
  Z.\,K. Liu,  S.\,K. Mo,  X.\,L. Qi,
  H.\,J. Zhang,  D.\,H. Lu,  X.~Dai,
  Z.~Fang,  S.\,C. Zhang,  I.\,R. Fisher,
  Z.~Hussain,  and  Z.\,X. Shen,
 Science \textbf{325}(5937), 178 (2009).


\bibitem{CAS2012}% article
 {F.~Casper},  {T.~Graf},  {S.~Chadov},
  {B.~Balke},  and  {C.~Felser},
 {Semicond. Sci. Technol.} \textbf{27}, 063001 (2012).


\bibitem{HAA02}% article
 {M.\,G. Haase},  {T.~Schmidt},  {C.\,G. Richter},
  {H.~Block},  and  {W.~Jeitschko},
 {J. Solid State Chem.} \textbf{168}, 18 (2002).


\bibitem{CAS08}% article
 {F.~Casper} and  {C.~Felser},
 {Z. Anorg. Allg. Chem.} \textbf{634}, 2418 (2008).


\bibitem{Chadov2010}% article
 {S.~Chadov},  {X.~Qi},  {J.~K\"ubler},  {G.\,H.
  Fecher},  {C.~Felser},  and  {S.\,C. Zhang},
 {Nat Mater} \textbf{9}(7), 541--545 (2010).


\bibitem{OUA2011}% article
 {S.~Ouardi},  {G.~Fecher},  {C.~Felser},
  {J.~Hamrle},  {K.~Postava},  and  {J.~Pistora},
 {Appl Phys Lett} \textbf{99}, 211904 (2011).


\bibitem{OUA2011b}% article
 {S.~Ouardi},  {C.~Shekhar},  {G.~Fecher},
  {X.~Kozina},  {G.~Stryganyuk},  {C.~Felser},
  {S.~Ueda},  and  {K.~Kobayashi},
 {Appl Phys Lett} \textbf{98}, 211901 (2011).


\bibitem{SHA12}% article
 {R.~Shan},  {S.~Ouardi},  {G.~Fecher},  {L.~Gao},
  {K.~Roche},  {M.~Samant},  {C.~Vidal~Barbosa},
  {E.~Ikenaga},  {C.~Felser},  and  {S.~Parkin},
 {arXiv:1209.5710 [cond-mat.mtrl-sci]} (2012).

\bibitem{SHE12}% article
 {C.~Shekhar}, {S.~Ouardi}, {G.~Fecher},
  {A.~Nayak},  {C.~Felser},  {E.~Ikenaga},
 {Appl Phys Lett} \textbf{100}, 252109 (2012).

\bibitem{MIY2012}% article
 {T.~Miyawaki},  {N.~Sugimoto},  {N.~Fukatani},
  {T.~Yoshihara},  {K.~Ueda},  {N.~Tanaka},  and
  {H.~Asano},
 J. Appl. Phys. \textbf{111}, 07E327 (2012).


\bibitem{Oestreich2003}% article
 J.~Oestreich,  U.~Probst,  F.~Richardt, and
  E.~Bucher,
 J. Phys.: Condens. Matter \textbf{15}, 635 (2003).

\bibitem{SHA12b}% article
 {R.~Shan},  {S.~Ouardi},  {G.~Fecher},  {L.~Gao},
  {A.~Kellock},  {A.~Gloskowskij},  {C.~Vidal~Barbosa},
  {E.~Ikenaga},  {C.~Felser},  and  {S.~Parkin},
 {arXiv:1209.5707 [cond-mat.mtrl-sci]} (2012).

\bibitem{Xia2001}% article
 {Y.~Xia},  {V.~Ponnambalam},  {S.~Bhattacharya},
  {A.\,L. Pope},  {S.\,J. Poon},  and  {T.\,M.
  Tritt},
 {J. Phys.: Condens. Matter} \textbf{13}, 77 (2001).


\bibitem{Uher1999}% article
 {C.~Uher},  {J.~Yang},  {S.~Hu},  {D.\,T.
  Morelli},  and  {G.\,P. Meisner}\iffalse Transport properties of pure
  and doped mnisn (m=zr, hf)\fi,
 {Phys. Rev. B} \textbf{59}, 8615 (1999).


\bibitem{Ouardi2010}% article
 {S.~Ouardi},  {G.\,H. Fecher},  {B.~Balke},
  {X.~Kozina},  {G.~Stryganyuk},  and  {C.~Felser}\iffalse
  Electronic transport properties of electron- and hole-doped semiconducting
  c1b heusler compounds: Niti1?xmxsn (m=sc, v)\fi,
 {Phys. Rev. B} \textbf{82}, 085108 (2010).


\bibitem{Ouardi2011}% article
 {S.~Ouardi},  {G.\,H. Fecher},  {C.~Felser},
  {J.~Hamrle},  {K.~Postava},  and  {J.~Pistora}\iffalse
  Transport and optical properties of the gapless heusler compound ptysb\fi,
 {Appl Phys Lett} \textbf{99}(21), 211904 (2011).


\end{thebibliography}
%
% Replace the following example bibliography with your references
% before submission:

%\bibitem{sales96}% article

%\end{thebibliography}

\end{document}